# Privacy Dashboards for Citizens and corresponding GDPR Services for Small Data Holders: A Literature Review *

Nico Puhlmann[1], Alex Wiesmaier[2], Patrick Weber[2], and Andreas Heinemann[2]

[1] PRODYNA SE, Neu-Isenburg, Germany
[2] Hochschule Darmstadt – University of Applied Sciences, Germany

**Abstract.** Citizens have gained many rights with the GDPR, e.g. the right to get a copy of their personal data. In practice, however, this is fraught with problems for citizens and small data holders. We present a literature review on solutions promising relief in the form of privacy dashboards for citizens and GDPR services for small data holders. Covered topics are analyzed, categorized and compared. This is ought to be a step towards both enabling citizens to exercise their GDPR rights and supporting small data holders to comply with their GDPR duties.

## 1 Introduction

### 1.1 Context

The European General Data Protection Regulation (GDPR) is one of the most stringent privacy laws world wide. It was presented in 2014, passed by the EU in 2016, and came into force in 2018. It affects everyone processing personal data of European citizens, even if the processor is not located inside the EU. It grants citizens several rights such as transparency, rectification, and the right to be forgotten. Although most people are aware of privacy problems of Internet-based platforms and services and indicate in studies that the protection of their privacy is important to them, they behave the opposite way [65].

### 1.2 Problem

In practice, it is a challenge for citizens to execute their rights granted to them by the GDPR [19]. In our opinion, the major reasons for that are that they usually: (C1) do not know who holds personal data of them, i.e. who they need to contact. (C2) need to contact the data holders one by one. (C3) are presented

* This research work has been partly funded by the German Federal Ministry of Education and Research and the Hessian State Ministry for Higher Education, Research and the Arts within their joint support of the National Research Center for Applied Cyber-Security ATHENE.

with a cacophony of interfaces and data formats by the different data holders. (C4) are not aware of the information that has been deduced from their personal data by the data holders.

On the other hand, small data holders (SDHs) such as SMEs and clubs might face problems complying with the regulation. In our opinion, the major reasons for that are that they often: (D1) lack the respective knowledge or skills. (D2) lack proper support to comply with their GDPR duties. (D3) cannot afford to pay for respective consulting.

### 1.3   Approach and Contribution

An approach to ease the issues stated in Sec. 1.2 is the provisioning of proper and affordable tooling. The work at hand presents the results of our systematic literature review (SLR) on privacy dashboards for citizens and GDPR services for SDHs. We show to what degree these are represented in current research, what features and solutions are proposed, which trade-offs they entail, and which issues remain open. We do that along defined research questions (Sec 3.2) focusing on functionalities, architectures, challenges, open research questions, and solutions.

### 1.4   Methodology and Structure

We start by introducing some background information and related work in Sec. 2. Then, in Sec. 3, we describe the methodology used to conduct the SLR. After that, the results are presented, subdivided into results for privacy dashboards in Sec. 4 and results for SDH services in Sec. 5. An evaluation of the results in the light of the research questions is given in Sec. 6, followed by a conclusion in Sec. 7.

## 2   Background

### 2.1   Terminology

We present the definitions of the major terms used within the work at hand. Where possible we use suitable definitions from literature, for others we give our own definitions.

**Data Subject** "A natural person (i.e. not a company or organisation) who resides in the EU, whose personal data is being processed by a controller."[3]
**(Data) Controller** "Simplified the controller is the individual or legal person who determines the purposes for which and the means by which personal data is processed."[1]
**Data Holder** same as Data Controller
**Personal Data** "Simplified it is the data relating to a physical person who with this data can be identified directly or indirectly."[2]
**Transparency** Forms of information visibility, as specified by Turilli et al. [72].

**Citizen** A native or naturalized person who owes allegiance to a government and is entitled to protection from it.

**Small and medium-sized enterprise (SME)** A medium-sized company with fewer than 250 employees and financial indicators such as turnover not greater than €50 million or balance sheet total not greater than €43 million [4].

**Small data holder (SDH)** Small organizations such as SMEs or associations and clubs interacting with a number of data subjects in the low hundrets.

## 2.2   Related Work

In [50] Murmann et al. present tools to achieve usable ex-post transparency, which includes privacy dashboards. While some of the work presented in their paper can also be found in the paper at hand, the paper at hand is different in scope. While Murmann et al. present literature about ex-post transparency tools, the paper at hand reviews literature about privacy dashboards. Also, Murmann et al. do not have a specific focus on SDHs that are explicitly considered in the paper at hand. While the work of Murmann et al. has been presented in 2017, in the paper at hand we also cover newer work that has been published since then.

In [41] Johannsen and Kant present a study that includes interviews and an examination of existing tools as well as a literature review. They examine how SMEs cope with IT governance, risk and compliance management. They conclude that there is no need for general approaches, but that distinctions must be made between industries sectors and company-specific circumstances. They also suggest that a distinction should be made between small and micro enterprises when developing new tools and standards.

In [76] Waidelich and Schuster investigate challenges that the GDPR poses to SMEs and their respective solutions. They propose a set of privacy patterns for realizing legal requirements within technology design. Their ultimate goal, which is subject to future work, is to provide a validated pattern catalog, possibly as set of method cards, and provide it to the public.

In [61] present a study on existing literature focusing on challenges posed by GDPR to business enterprises and outline future research. They restrict themselves on Scopus and Web of Science databases. They find that existing literature is often limited to certain countries and technology-focused sectors. The also state that SME's are more often in the focus of GDPR research than larger enterprises.

In [44] we present a survey of existing GDPR solutions for citizens and SMEs from a market point of view. The paper at hand complements this by investigating the respective state of research.

## 3   Planning and Conducting the SLR

The SLR is performed based on the guidelines provided by Kitchenham and Charters [43]. Our SLR comprises the following steps that are detailed in the subsequent sections:

1. Definition of the formal scope of the SLR.
2. Definition of the research questions.
3. Definition of the search strategy.
4. Definition of the search terms.
5. Definition of the literature selection criteria.
6. Definition of the data extraction method.

### 3.1   Formal Scope

The formal scope of the review builds upon the Taxonomy of Literature Reviews by Randolph [53]. The taxonomy is structured in six characteristics that allow to classify literature reviews. Each characteristic is split into one or more categories that specify how the review is positioned in the respective context. In the following, the classification for the paper at hand is given, according to the mentioned taxonomy. For each of the six characteristics, we name the respective categories (in italics) that apply to our review.

**Focus** We focus on the *research outcomes*, *theories* and *practices or applications* of the reviewed papers in the given context and related to the research questions.

**Goal** We aim to provide an *integration* of the solutions, found in the reviewed publications. Also, the *identification of central issues* is a goal of the paper at hand.

**Perspective** During the review, a *neutral representation* of the reviewed work is intended.

**Coverage** The review is designed to be *exhaustive with selective citations*. Exhaustive in this context means, that it considers all available publications from the selected sources. It is selective in that the choice of the databases and search terms may be considered selective. We aim to cover a representative selection of work in the context of computer science and engineering.

**Organization** The paper at hand is organized in *conceptual format* and in *methodological format*. It is methodological in the sense, that it is structured like a scientific paper, presenting an introduction, method, results and discussion. It is structured conceptual, as the reviewed publications are structured by the topics they cover.

**Audience** The focused audience of the paper at hand are *specialized and general scholars* in the field of information science. It is also meant to serve as an overview for *practitioners*, who want to develop a privacy dashboard or service for SDHs.

### 3.2 Research Questions

The research questions were defined during a pre-study phase, in which the authors of the paper at hand discussed the topic. The overarching research question (RQ0) and its sub-questions (RQ1) - (RQ6) are:

- (RQ0) What is the current state of research in privacy dashboards for citizens and GDPR services for SDHs?

  - (RQ1) Which functionalities of privacy dashboards for citizens are proposed in literature?
  - (RQ2) Which functionalities of GDPR services for SDHs are proposed in literature?
  - (RQ3) Which architectures are proposed for said dashboards or services?
  - (RQ4) What are the challenges in providing said dashboards or services?
  - (RQ5) Which open research questions in this context are formulated or can be derived?
  - (RQ6) Can we derive solutions from given approaches found in literature?

### 3.3 Search Strategy

The search was limited to abstracts, title and keywords, to exclude hits that might only mention a search term once in the paper. Different scientific sources were used, to reduce the search selection bias. The sources used are: ACM Digital Library, Science Direct, IEEE Explore and Google Scholar. Springer was considered as an additional database, but does not allow to limit the search terms to abstract, title and keywords. Nevertheless, Springer publications were found through Google Scholar and considered in the work at hand. We considered only literature that was published from 2014 as this was the year when the EU parliament adopted the GDPR. Some databases still yielded too many results, due to the widely used terms such as DSGVO and dashboard. In such cases, The first 200 results per database were checked for relevance in our context. The initial search was conducted in February 2022. Update searches were conducted in September 2022 and March 2024[3]. Compared to the previous searches, the last search was conducted using a broader set of search terms that are given in Sec. 3.4. Notably, the broader search, which just added more OR connected terms to the original query, did not find one of the matching papers we found using the smaller query in previous searches. We kept this paper [46] in the survey, and also added some papers [16,77] that did not result from any of the searches, but were mentioned to us by colleagues.

For each database the search was done in multiple phases:

**Phase 1** The search query was executed against the current database
**Phase 2** The abstracts of newly found literature were read
**Phase 3** Literature matching our context was read further
**Phase 4** A forward and a backward search was conducted, for resulting papers Phase 2 and Phase 3 were performed

---
[3] March 10th to March 12th 2024

### 3.4  Search Terms

In a first step, the initial search terms were extracted from the main research question RQ0. These are: (1) General Data Protection Regulation (2) privacy dashboard (3) small data holders. In a ramp-up phase, the initial search terms were used to identify alternative terms within related papers and the Internet and by discussions in the team. As the authors speak German, it was decided to additionally use the German equivalents to the English search terms, as this at most broadens the search results. As the interest was in two different thematic domains ($d_1$) a GDPR dashboard for citizens ($d_2$) a GDPR service for SDHs, this resulted in two distinct sets of search terms.

The domain $d_1$ for privacy dashboards represents the conjunction of the subdomains GDPR and privacy dashboards:

$$d_1 := GDPR \wedge Privacy\ Dashboard \tag{1}$$

Relevant publications must therefore be located at the intersection of these two areas. For the domain $d_2$ of services for SDHs, it is the conjunction of the two subdomains GDPR and SDHs:

$$d_2 := GDPR \wedge Small\ Data\ Holders \tag{2}$$

The publications must reside in the context of the given domains. Thus, the following search terms are defined:

$$\begin{aligned}
GDPR := \\
& GDPR \vee DSGVO \vee right\ of\ access \vee \\
& Recht\ auf\ Auskunft \vee priva* \vee transpar* \vee \\
& data\ rights \vee data\ protection \vee Datenrecht \vee \\
& Datenschutz
\end{aligned} \tag{3}$$

$$\begin{aligned}
Privacy\ Dashboard := \\
& Privacy\ Dashboard \vee Dashboard \vee cockpit \vee \\
& citizen \vee Bürger
\end{aligned} \tag{4}$$

$$\begin{aligned}
Small\ Data\ Holders := \\
& SME \vee KMU \vee club \vee Verein \vee \\
& small\ data\ holders \vee kleine\ Datenhalter \vee \\
& association \vee small\ business \vee small\ company \vee \\
& Kleinunternehmen \vee Kleinbetrieb
\end{aligned} \tag{5}$$

The term *GDPR* is augmented by several contextually related terms (Term 3). Its German translation is *DSGVO*. The *right of access* is anchored in the GDPR and describes the legal claim of individuals to get information on their personal data held by others. It is an integral part of the context of privacy dashboards. The German translation of the term is *Recht auf Auskunft*. Privacy describes the concept that individuals have sovereignty over their personal data, which is the central purpose of the GDPR. The respective German translations *Privatheit* respectively *Privatsphäre* share its beginning *priva*, so a wildcard expression is used here to include all of them. Transparency describes the principle that all relevant information on the collection and use of personal data is concisely and clearly communicated to the data subject. Again, a wildcard expression starting with *transpar* is used to include the German translation *Transparenz*. Finally, the terms *data rights* and *data protection* are considerend, together with their respective German translations *Datenrecht* and *Datenschutz*.

The term *Privacy Dashboard* (Term 4) is augmented by the more general terms *Dashboard* and *cockpit* that may also be used within GDPR related texts to denote a Privacy Dashboard. As these English terms are commonly also used in German, no translation needs to be added. Finally, the term *citizen*, and ist German translation *Bürger*, establishes the link to the target group.

The term *Small Data Holders* (Term 5) is defined by the acronyms for small and medium enterprises, i.e. *SME* in English and *KMU* in German. This is augmented by *club* and its German translation *Verein*. This is augmented by terms with similar but more specific meanings, where *kleine Datenhalter* is the German term for *small data holders*, *Kleinunternehmen* is the German term for *small business* and *Kleinbetrieb* is the German term for *small company*. The German term for *association* is *Verein*, which is already included.

During the pilot search, it was found, that the acronyms like GDPR or SME are widely used and for scientific literature often already are used in the abstract and even in the title. Therefore, the spelled out words were not used in the search.

### 3.5 Literature Selection Criteria

The literature had to match the following selection criteria:

- Language in English or German
- Full text access available through account of Darmstadt University of Applied Sciences. [4]
- Context matches thematic domains $d_1$ or $d_2$ (cf. Sec. 3.4).

### 3.6 Data Extraction Method

During the SLR, all resulting papers were matched against the study selection criteria, defined in Sec. 3.5 and the scope defined in Sec. 3.1. Matching papers were than read in the following order with early abortion: Abstract, Conclusion, Discussion, Introduction, Main Area. Most of the publications found with the used queries, were found not to be relevant for the paper at hand. Some publications were written in languages other than German or English (e.g. because of English technical terms) and therefore excluded.

---

[4] Access to the major publishers is given.

Table 1 shows the total number of papers found for domain $d_1$ per database together with the number of papers out of this set that were eventually cited within the paper at hand. Table 2 shows the same for domain $d_2$. Due to size restrictions of the Science Direct search interface, we had to break up the search query into multiple smaller queries. The tables show the totals of the individual queries, separated by colons.

**Table 1.** Total number of papers found and cited for search domain $d_1$

| Database | Total Found | Papers Cited |
|---|---|---|
| ACM | 3070 | 1 |
| IEEE Explore | 2010 | 2 |
| Science Direct | 42 : 637 : 0 | 2 |
| Google Scholar | 25800 | 34 |

**Table 2.** Total number of papers found and cited for search domain $d_2$

| Database | Total Found | Papers Cited |
|---|---|---|
| ACM | 7 | 0 |
| IEEE Explore | 2427 | 1 |
| Science Direct | 17 : 308 : 3 : 50 : 411 : 8 | 0 |
| Google Scholar | 25800 | 10 |

The forward and backward search was conducted using Citation Gecko.[5] During the initial search, the tool provided one publication of interest in forward direction and 6 in backward direction. There were no additional papers found for the last update search.

## 4  Dashboards

In total there were 46 publications, dealing with privacy dashboards.

### 4.1  Topics, Attention, and Scope

This subsection presents the topics, the publication timeline, and the scope covered by the different publications on privacy dashboards found during the literature review.

---

[5] https://citationgecko.azurewebsites.net/

**Publications by Topic** As shown in Tab. 3, the field of privacy dashboards is divided into several individual topics, e.g. different designs can be investigated or requirements management can be performed. A publication can cover several subtopics, e.g. a design proposal and a user study. Each publication is assigned to its main topics (one or more). For each topic, the number of assigned publications is counted to find the most frequently mentioned topics. Within the total number of 46 publications, the fields of usability, user studies and architecture are the most covered, with 16 publications for usability, 15 for user studies and 12 for architectures, while Launch, i.e. the roll out of dashboards is represented in three, and Requirements Engineering (RE) in four works. The remaining 18 publications each deal with individual topics and have been collected under *Other*. The high number of individual other topics indicates that more research needs to be done at least in these areas of privacy dashboards.

Table 3. Publications about privacy dashboards and their (main) topics.

| Reference | Usability | User Study | Architecture | Launch | Requirements Engineering | Other |
|---|---|---|---|---|---|---|
| [17] | | | | | | Presentation of EU funded projects about privacy dashboards |
| [63] | | | | | | Research on existing Transparency Enhancing Tools (TETs) |
| [62] | | | | X | | |
| [68] | X | | | | | |
| [70]. | X | | | | | Comparison of data exports and privacy statement vs. dashboards of existing solutions |
| [15] | X | | | | | Framework for implementation of privacy dashboard in corporate environments |
| [30] | | | | | | Crowdsourcing platform and browser plugin |
| [34] | X | X | | | | |
| [24] | | X | | | | User sentiment after usage of privacy dashboard |
| [32] | | | | | | Overlooked problems of privacy literacy |
| [45] | | | | | | Overlooked problems of privacy literacy |
| [48] | | | | | | Use cases of privacy dashboard in industry 4.0 + requirements |
| [7] | | | | | | Privacy dashboard protection principles in health care environment |
| [80] | | | | | | Classification scheme for privacy dashboards |
| [54] | X | | X | | | |
| [5] | X | X | | | | |
| [42] | X | X | | | | |
| [74]: | X | X | | | | |
| [64] | | | | | | Exposure to third parties in Facebook environment |
| [60] | X | X | | | | |
| [59] | X | X | X | | | Unification scheme |
| [75] | X | X | | | | Data curation |
| [49] | X | | | | | |
| [28] | | | | X | | |
| [71] | | | X | | | |
| [20] | | | X | | | |
| [57] | | | X | | | |
| [10] | | | X | | | |
| [9] | | X | X | | | |
| [56] | | | X | | | |
| [78] | | | X | | | |
| [13] | | | X | | | |
| [52] | | | | X | | |
| [21] | | | | | X | |
| [14] | | | | | X | |
| [67] | | | | | X | |
| [69] | X | X | | | | |
| [29] | | | X | | | |
| [26] | | | | | | Research on existing Transparency Enhancing Tools (TETs) |
| [55] | | X | | | | IoT Framework including dashboard for data subjects |
| [66] | | X | | | | User study on privacy notifications not limited to dashboards |
| [8] | | X | | | | |
| [25] | | X | | | | |
| [38] | X | | | | | |
| [77] | X | X | | | X | |
| [16] | X | X | | | | |
| Total: 46 | 16 | 15 | 12 | 3 | 4 | 18 |

**Publications per Year** Fig. 1 shows that the number of publications has increased significantly in 2020 and 2021 and dropped again in 2022 and 2023. This shows that the topic of privacy dashboards gained momentum with a delay of two years after the GDPR came into force, but dropped significantly after only two more years. At the time of writing the paper at hand, in March 2024, there was no paper from 2024 found on the topic.

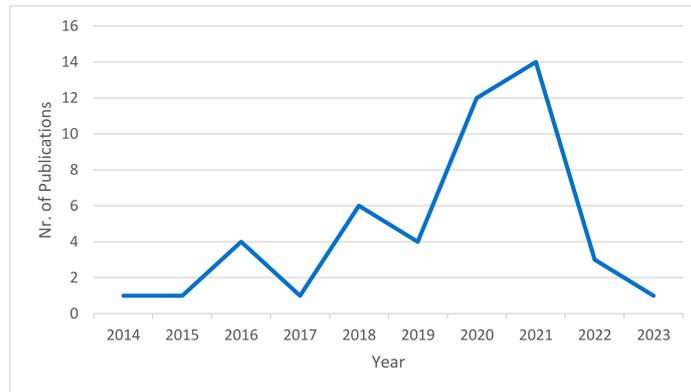

**Fig. 1.** Number of publications on privacy dashboards per year between 2014 and 2023.

**Scope of the Dashboards** Fig. 2 shows the proportions of dashboards that are intended to display data of multiple data holders (10) compared those that are intended to display data of exactly one data holder (27). Thereby it does not matter whether a dashboard is installed locally or used as a web service. The share of dashboards that are only for one data holder is further subdivided into dashboards that serve internal purposes (5), e.g. to give employees access to their data, and those that are to be used for external purposes (22), e.g. to be made available to customers. It can be seen that the number of proposed dashboards for multiple data holders is almost a third the number of dashboards for only one data holder. Also, within the dashboards for only one data holder, the number of dashboards that are meant to serve external purposes, is more than 4 times higher than the number of dashboards serving internal purposes. Despite the strict differentiation of internal and external dashboard purposes in the literature, the technical difference between internal and external dashboards remains unclear.

Fig. 3 shows the target environment of the proposed dashboards. Here, it is not considered whether the dashboards are for internal or external purposes, but in which application context they are to be used. We see that the proportion of single target (22) dashboards outweighs the multi purpose dashboards (15), but they are still relatively close.

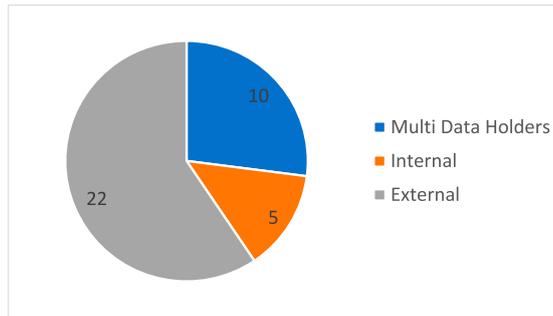

**Fig. 2.** Single vs. multi data holder dashboards within the literature.

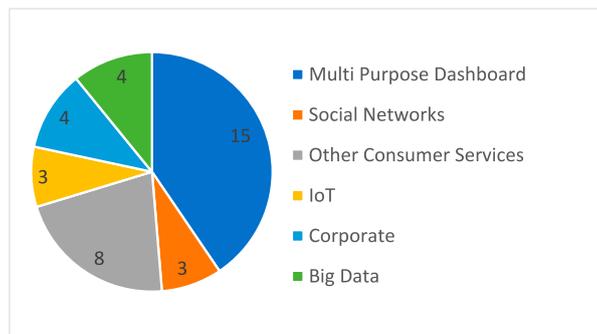

**Fig. 3.** Proportion of target environments within the literature on privacy dashboards.

### 4.2  Approaches and Technologies

**Existing Solutions** The publications that deal with existing Transparency Enhancing Tools (TETs) and Privacy Enhancing Tools (PETs) have found them to be inadequate. The reasons for this vary. For example, [68] argue that some of the tools had only a small range of functions and only covered special scenarios. As shown by [70] some dashboards do not display all the information that is supposed to be provided, for example technical data such as the IP address or the operating system used.

Studies that look at data subjects' trust towards a data holder show that trust can be strengthened by a privacy dashboard [24,34]. Trust in the data holder actually becomes the stronger the more control data subjects have over their data [34]. Another study [24] shows that perceived risk and concern about one's online behavior is lower when using a privacy dashboard.

But, this gain in trust and perceived control also may do more harm than good. According to [32], persons well trained in the use of computers, are more likely to feel in control of the situation and therefore disclose more data. For this reason, [45] argue that the responsibility for privacy should not be left to the data subject, but should be enforced through stricter laws. They argue that education cannot take place on the scale it is needed. However, they also write that the laws must be well-considered and the balance between individual autonomy, consumer protection and restriction by the state must be ensured.

Another problem is that while the data collected by a company is made available to the data subject, information what data has been given to third parties is usually not included [64]. Thus, data subjects often do not know which data is transferred by who to whom and therefore cannot make any GDPR requests to the respective data holders.

**Usability** User studies have shown that data subjects are not necessarily surprised about the data stored by different providers [42,74,5,60].

The publications reviewed make slightly different suggestions for the design of a privacy dashboard. But, they all have in common that they propose to present data in different layers, a simplified overview on top and more details below. In [60] the authors propose using artificial intelligence and a human in the loop approach to find the most important data and present it in the overview. This overview shall be designed to cope with growing volumes of data, to not overwhelm the data subjects with increasing data over time.

The user interface should support filtering and searching and it should be interactive, e.g. providing a zoom function for more detailed data [74].

Respondents in various studies have also shown interest in derived data e.g. [42,5,74]. Showing data subjects what information could be derived from existing data would therefore be a step toward a better understanding of data privacy.

The available data exports are too complicated for data subjects to understand, as many still contain internal designations [74]. It is thus not clear from the data what exactly they represent. A user interface should therefore designate the data in a comprehensible way. For software that visualizes data exports, this would mean that the exported data would have to be "translated" for each data holder.

**User Study** User studies are a main topic according to the previous analysis, but they were, with the exception of [66] and [8], conducted on at least one other main topic. These studies are not discussed here but considered together with their respective main topic.

**Architectures** When it comes to privacy dashboard architecture publications (cf. Tab. 3) basically three different approaches regarding the dashboard hosting are visible:

(1) The dashboard is hosted by the data holder.
(2) The dashboard is hosted locally by the data subject.
(3) The dashboard is hosted by a trusted third party.

Further, we identify two approaches regarding the provisioning of the dashboard:

(A) The dashboard is provided by the data holder.
(B) The dashboard is provided by a trusted third party.

The predominant approach is for data holders to provide a dashboard themselves (A), which is then either hosted by the data holder (1), or on the local computer of the data subject (2).

Fig. 4 shows the distribution of architectures across the publications covered. Not only actual proposals for architectures are considered, but also the proposal of an approach. For example, [54] proposes to implement the dashboard locally and connect to all data holders, but does not specify how this should be implemented. We see that only two of the proposed dashboards are to be run on the local computer of the data subject. Six propose the dashboard to be run at a trusted third party, while 20 propose it to be run at the data holder itself.

**Launch of the Dashboard** All the publications that have been studied and propose the introduction of a dashboard suggest doing so gradually [21,14,67]. The authors of [14] suggest that the functions required by regulation should be introduced first. Further steps could then introduce rights control and other more advanced views. [67] provide suggest a five step plan to introduce privacy dashboards. In the first step, it should be possible to obtain information on the processing of personal data. In further expansion stages, requests for consent to the collection of data, the embedding in existing systems and the enforcement of rules and regulations will be introduced. [21]'s work deals with the introduction of a portal for citizens to view pension information. The introduction of the portal is to take place in several phases and will first connect all pension institutions that are required by law to provide regular information. A step-by-step connection of further information sources is recommended by the authors.

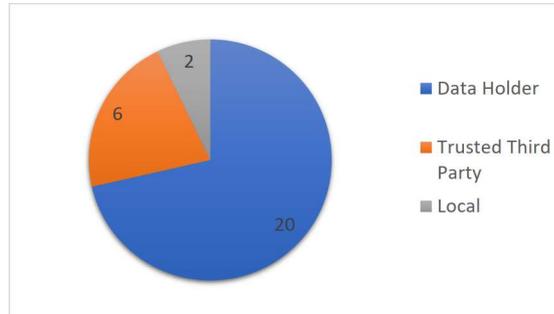

**Fig. 4.** Distribution of proposed architectures across publications.

### 4.3   Open Issues and Unresolved Questions

From Tab. 3, one can see that there are many papers on other topics besides the main topics. The share of papers sorted into the category *Others* in the total number of publications found is just about 39%. In the following, a selection of the publications summarized under *Other* are discussed in order to reveal open questions and problems.

In [70] Tolsdorf et al. examine the response to the execution of the right of access of different data holders. The study reveals that for many data holders, essential information is not made available. This includes technical information such as the IP address, but also information about third parties who may have access to the personal data. The question remains open as to whether the data holders themselves do not have this information, or if it is simply not made available. The time between request and provisioning of data also varies, sometimes significantly, between data holders. For example, Google's data was available within 30 minutes, while Amazon's data was only ready for download after four days. The comparison between data exports and the privacy dashboards offered showed that the data exports contained between 14% and 22% more data than was available in the respective privacy dashboard. This shows that the currently available privacy dashboards are not (yet?) fully functional. Data subjects currently have no way to verify that a privacy dashboard displays all available data collected by the data holder.

When it comes to dashboard architecture, current implementations are mostly done directly on the site of the data holder. This is often a problem for SDHs, as there is a lack of technical knowledge, money and personnel to implement this. Therefore, Raschke et al. [54] prefer an architecture where a dashboard connects many data holders and serves as a single point of contact for data subjects. However, this is difficult to implement due to the many different ways data is stored, different formats, and the sensitive nature of such a service. The only technical proposal to implement such a service is therefore based on the data subject continuing to request his data individually from each data holder and then analyzing the data exports in a locally executed program [59]. This approach continues to be a challenge for the developers of such solutions, as the exported data is stored in different formats. Also, the individual information may be named differently by different data owners and internal designations may not be interpreted at all. Such a solution therefore relies on an individual implementation of the preparation of data per different data holder. Such an implementation is prone to changes in formats and designations by the data owner and therefore requires continuous maintenance of the individual components. This solution would also have to be executable on multiple platforms and operating systems. This was solved by Schufrin et al. [59] by running the program in the browser. However, this could lead to problems with regard to performance for large amounts of data.

One problem that is, for example, addressed by [29] is that data subjects do not know who is storing their data in the first place. Due to the sheer number of services available, it is, at least, difficult to get an overview. Even if a data subject knows all his data holders, he can't be sure to whom else they have passed the data. The solution proposed by [29] requires data holders to collaborate and use the same system, which might raise new problems.

Two studies [45,32] raise the question of whether data subjects have or even can build up the competencies to regulate the protection of their own data themselves. It is particularly difficult for data subjects to recognize data protection settings, processing of data on the backend site, and the transfer of data to third parties. Current approaches to simplify the data protection settings and introduce privacy symbols do not go far enough for the authors, they argue that regulation should be left to the legislature. This allows for a cross-topic question of whether a privacy dashboard can actively help data subjects take responsibility for their privacy and whether using a dashboard encourages data subjects to be more mindful of their personal data.

The study of Farke et al. [24], shows that data subjects had fewer concerns about handling their personal data after using a privacy dashboard. However, said study is limited to only one dashboard and 153 data subjects, and the impact of other dashboards and dashboard design on data subjects was not examined. The approach of using AI to present the most sensitive data of a data subject at a glance [60] could possibly cause different reactions of data subjects.

## 5 Services for SDHs

### 5.1 Topics and Attention

**Publications by Topic** As seen in Tab. 4, the field of publications dealing with GDPR services for SDHs is not very wide. It can be seen that ten (cf. Sec. 5.1) out of 18 publications found, deal with business process management. In general, they deal with changing the processes within a company so that the company meets the requirements of the GDPR. The remaining eight publications each deal with their own topics and are therefore grouped under *Other*.

**Table 4.** Publications about solutions for data holders and their (main) topics.

| Reference | Business Process Management | Other |
|---|---|---|
| [11] | | Empirical research on data protection practices in SMEs |
| [17] | | Presentation of EU funded projects |
| [40] | X | |
| [47] | X | |
| [6] | X | |
| [58] | X | |
| [33] | X | |
| [23] | X | |
| [35] | X | |
| [51] | | Integration of existing tools |
| [22] | | SMOOTH Project |
| [12] | | Authentication of GDPR requests |
| [31] | | Transparency language scheme and document store |
| [46] | X | |
| [37] | | Architecture for implementation of GDPR |
| [18] | X | |
| [27] | | RuleKeeper Project |
| [79] | X | |
| Total: 18 | 10 | 8 |

**Publications per Year** Fig. 5 shows that the number of publications peaked in 2019 and 2020. The first publication [11] is from 2015, however does not cover solutions for SDHs, but shows that the number of requests for own data grows with the number of employees of the company. On average, small businesses receive far fewer requests than large companies. However, the investigation was conducted three years before the GDPR came into force and may not be representative for today's situation.

As with the dashboards (cf. Sec. 4.1), we see a decreasing number of publications after a short peak. At the time of writing the paper at hand, in March 2024, we found two publications from 2024 on the topic.

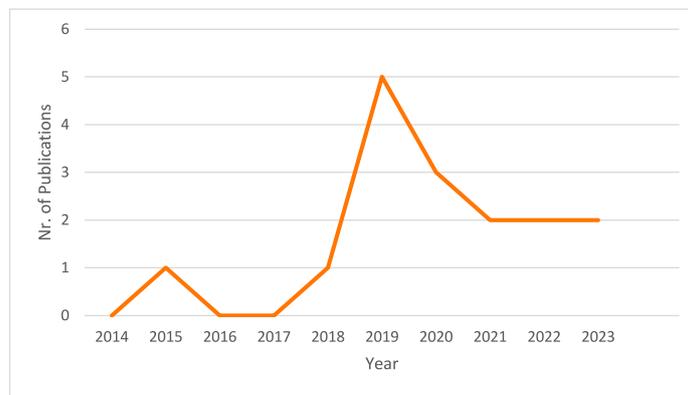

**Fig. 5.** Number of publications about SDH solutions per year between 2014 and 2023.

### 5.2 Approaches and Technologies

**Process Management** Ten publications suggest processes or process changes to make companies GDPR compliant. We look at the various methods that are used to achieve this change. The authors of [47] propose iterative procedures in which different stages are passed through in a cycle. [6] suggests a waterfall model like approach in which each previous step is completed in order to begin the next. A suggestion presented in [58] is to use enterprise architecture frameworks, as these generate an overview of the company and can thus also show data flows. One project, presented in [23], has already partially automated its procedure so that companies have less effort in going through the process.

Despite the different ways to achieve their goal, in essence all publications on process management use the following approach. The first step is for the company to get an overview of the personal data it processes. Here, the data of employees must also be explicitly taken into account, as they can exercise the same rights with regard to the GDPR as a regular data subject, such like customers. The next step is for the company to consider how the data is collected, how long it is stored, for what purpose it is needed, et cetera. Technical information should also be taken into account. For example, if the data is stored using a database management system, there are possibly backups of this data. In this case, if deleting the data, care must be taken to ensure that the data is also deleted from the backups. In the final step, the company considers new processes to meet the requirements of the GDPR. These processes are then integrated into the company's work processes. These approaches require good documentation of the results in each step. Optionally, internal evaluations can be conducted with stakeholders and responsible parties.

**Other** Three of the remaining papers propose technical solutions for SDHs. In [51], an orchestrator platform is proposed that manages and controls individual and existing software to support the implementation of the GDPR. Another publication proposing a technical solution is introduced in a paper that presents EU founded projects [17]. The SMOOTH project is meant to help micro enterprises assess their GDPR compliance. It supports the upload of files such as consent forms and terms of use as well as databases containing personal data. As output, the micro enterprise receives a compliance report. In [31] the authors propose a technical solution that provides a way for data holders to store transparency information in a JSON scheme. The proposed document store is capable of being integrated into an existing infrastructure. However, it requires development effort on the side of the data holder and thus is not suitable for SHDs.

### 5.3   Open Issues and Unresolved Questions

There are only a few publications on solutions for GDPR request handling for SDHs. It remains unclear whether there is still a lot of room for further research or there is just no need for more research. Most publications in the field refer to business process management to make a company GDPR compliant. In [11] the authors show that only 4% of companies with less than 500 employees receive more than 10 GDPR requests per year. As this study dates back to 2015, three years before the GDPR came into force, further research should be conducted here, to see if this is still true. It should be clarified whether SDHs (including small local associations and micro-enterprises) are regularly confronted with GDPR requests and whether responding to them is a problem for them at all. It is also unclear which technology is usually applied to manage the personal data held by these SDHs and whether they can benefit from a service that helps them with GDPR requests. Not least, depending on the amount and type of personal data the SDHs collect, which needs to be investigated, a digital GDPR service may overshoot the mark.

Since the paper at hand is limited to scientific publications, it is not clear whether there are already technical solutions for SDHs on the market, that have not produced a scientific publication. The two existing publications [51,31] on technical solutions for SDHs could be an indicator that other technical solutions exist.

# 6  Discussion

In this section we discuss the state of research in privacy dashboards for citizens and GDPR services for SDHs from our findings so far. To this end, the remainder of this chapter is structured along the research questions from Sec. 3.2.

## 6.1  RQ1: Proposed Functions of Privacy Dashboards for Citizens

The functions that a privacy dashboard should demonstrate have been almost consistently defined in literature. Accordingly, a privacy dashboard should naturally present the data that a data holder has about a data subject e.g. [68,70]. Here, care should be taken to make it interactive and interesting for the data subject [74]. As many data subjects already know what data a data owner has about them, it is more interesting for them what further information can be derived [5,74,42]. A privacy dashboard should also be able to cope with the growing amounts of data that inevitably accumulate when a service is used for a long time. The privacy dashboard should be designed to first present an overview of the data, but also provide a way to analyze the data in detail. [74]

Most of the publications propose to implement the further rights that the GDPR offers to data subjects. The right to be forgotten and the right to rectification were frequently mentioned. Another proposed function was to compare the claimed data from privacy statements to the actual exported data from GDPR data exports [70]. Comparison of usage or volume of gathered data, between users of different social media networks has also been proposed [5].

To manage the risks of sharing data with third parties on social networks, e.g. by sharing contact lists, it was proposed [64] to extend a dashboard in such a way that a privacy score shows how high the risk is for a user to be affected by such sharing. However, this is not a trivial task, as it is an individual risk, depending on the behavior of third parties, e.g. the user's contacts.

A few papers [59,75] propose to provide a privacy dashboard for multiple data holders. The majority of publications deals with the implementation of dashboards at or for a specific data holder or for a specific purpose e.g. analysis of data from a particular social network. One publication [73] proposes to provide data subjects with a way to create an identity per data holder inside the privacy dashboard, allowing them to control what data the data subject discloses to the data holder. However, there is already a policy discussion in the EU about mandatory ID in social networks [36], so such an approach may not be feasible in the future.

A publication that proposes a dashboard that connects multiple data holders solves the problem of different data formats by using a unification scheme. A service per data holder transforms the received data into the new scheme so that the data can always be presented to the data subject in a uniform way. However, if one wants to implement this approach for each data holder, this comes along with a considerable development effort. If a data owner changes its format, the service has to be adapted, which requires continuous maintenance of the software.

### 6.2 RQ2: Proposed Functions of GDPR Services for SDHs

While only a few technical solutions are proposed in publications for SDHs, more publications [47,35,6,23,33,39,58] deal with the implementation and introduction of processes to ensure compliance with the GDPR. Similar procedures have been described for these processes. The respective procedures to reach this that are proposed by the different works are are similar, and can be summarized as follows. After identifying data sources, data stores and the surrounding processes, these are adapted to comply with the GDPR. The proposed technical solutions help either to check existing processes and documents for compliance with the GDPR or to orchestrate already existing tools. The proposed solution to store GDPR personal data information in a document store creates too much implementation effort for small businesses and is therefore not appropriate.

### 6.3 RQ3: Proposed Architectures for Privacy Dashboards and GDPR Services for SDHs

In terms of architectures of the proposed privacy dashboards, we identified three different approaches.

- Implementation per data holder
- Centralized implementation, connecting different data holders
- Local application on data subject side

The different approaches come with advantages and disadvantages that are discussed in more detail in Sec. 6.4. Many of the proposed privacy dashboards, regardless of where they are executed, propose microservice architectures where only one task is executed for each component of the system and it is delivered to the outside via APIs. No concrete architectures have been proposed for GDPR services for small data holders.

### 6.4 RQ4: Challenges in Providing Privacy Dashboards and GDPR Services for SDHs

Many of the publications have conducted user surveys and show that data subjects trust data owners more when they can manage their data via a privacy dashboard. However, these have also often not been implemented satisfactorily to date. Many of the dashboards do not show all data or are only for a specific service. Also, there are countervailing voices that show data subjects are less concerned about how their data is handled when they have had access to a privacy dashboard. This can give data subjects a false sense of security and tempt them to disclose more data that they might otherwise have kept to themselves. Some authors point out that data subjects should not be responsible for protecting their personal data themselves. They point to the complexity of today's technology and the fact that the necessary knowledge to understand it cannot be taken for granted.

Different data formats require a proper handling on both sides, the dashboard must be able to read and present the data from different data holders. So, work needs to be done on ways to unify the data formats, as has already been done in [59], but might also be investigated further.

### 6.5 RQ5: Open Research Questions

Many of the authors of the papers discussed here emphasize that they want to take their work further by extending their tools or conducting more user studies. Especially for SDHs there are few publications and the lack of solutions for SDHs in the scientific literature seems to be a relevant finding.However, the results of the study by Malinka et al. [11] also raise the question of whether SDHs are affected by GDPR requests at all. It is also unclear whether there are already solutions for SDHs , which were not found in the paper at hand because there is no scientific publication about it. The question of the data subjects' responsibility for their data has also been raised in some publications. The argument is that they cannot follow technical developments and are therefore unable to understand how and where their data is stored and processed. The question that remains open here is whether other laws or technical systems can provide a solution to this problem.

### 6.6 RQ6: Deriving a Solution

The complexity of the issues does currently not allow for deriving a solution directly from the publications. However, recommendations for developing solutions can be derived from the publications. For example, the concept for a privacy dashboard design can be derived from the summarized papers. Ultimately, however, it may not be possible to find purely technical solutions and it may be necessary to, e.g., adjust the laws and regulations to create a working GDPR ecosystem.

## 7  Conclusion

The paper at hand presents a literature review on privacy dashboards for citizens and GDPR services for SDHs. It addresses two interlinked groups: (1) citizens who want to execute their rights granted to them by the GDPR and (2) SDHs who need to provide the respective internal mechanisms and external interfaces.

Implementing (centralized) dashboards that are managed by third parties seems to be difficult. Not only the technical challenges resulting from the different data types and formats have to be mastered. Also, the question of who manages such a platform and whether this would be accepted on a voluntary basis by the data holders must be clarified. A platform or a system to which data subjects upload their data, which they have submitted to the data holders via a GDPR request, only partially solves the problems of the data subjects. They would still need to know who stores and processes their data, they would still not be able to verify that all data is available, and they would also need to learn how to use the program that prepares the data for them.

A question to be answered from a social science perspective in this case is, whether the responsibility for protecting data should lie with the data subject himself. Some believe this is no longer appropriate, as these explanations are often not comprehensible to laypersons and technical developments no longer allow ordinary data subjects to grasp how their data is processed. If this holds true, more intervention by the state and new regulations dealing with that may be appropriate. Since such legal changes usually take a long time, one solution in the meantime could be to provide data subjects with a service that guides them to a collection of their data holders and also shows them how to create a GDPR request with their data holders. There is also indication that SDHs do not need an automated GDPR service, as they are hardly affected by GDPR requests.

We saw that the current research in the field of GDPR services for SDHs hardly describes any technical solutions. In fact, our literature search reveals that there is a lack of publications in this area in general. In contrast, the field of privacy dashboards is already well studied, even if most of the proposed dashboards only integrate data from a single data holder. In particular, the usability and presentation of the data are well studied. There are different approaches in the literature on how such a dashboard should be provided. The most discussed approach is dashboards that are provided by a data holder to the data subjects. Followed by the approach in which a dashboard connects multiple data holders, creating a single point of contact for data subjects.

A next step in the research on the topic of privacy dashboards and GDPR services could be addressing specific uses cases or pilots to examine specific challenges in more detail.